\documentclass[12pt]{article}
\usepackage{palatino}
\usepackage{a4}
\usepackage{cite}
\usepackage{graphicx}
\usepackage{amssymb}
\usepackage{amsmath}
\usepackage{epsfig}
\usepackage[tmargin=1cm,bmargin=1cm,lmargin=2cm,rmargin=2cm]{geometry}

\newcommand{\qb}{\ensuremath{\overline q}}

\newcommand{\be}{\begin{equation}}
\newcommand{\bdm}{\begin{displaymath}}
\newcommand{\bea}{\begin{eqnarray}}
\newcommand{\beastar}{\begin{eqnarray*}}

\newcommand{\ee}{\end{equation}}

\newcommand{\edm}{\end{displaymath}}
\newcommand{\eea}{\end{eqnarray}}
\newcommand{\eeastar}{\end{eqnarray*}}

\newcommand{\rarrow}{\ensuremath{\rightarrow}}

\nopagebreak

\begin{document}
\baselineskip 22pt

\begin{flushright}
\end{flushright}
\vskip 65pt
\begin{center}
{\large \bf
Next-to-leading order QCD corrections to the $Z$ boson pair production
at the LHC in Randall Sundrum model.
} 
\\
\vspace{8mm}
{\bf
Neelima Agarwal$^a$
\footnote{neel1dph@gmail.com},
}
{\bf
V. Ravindran$^b$
\footnote{ravindra@hri.res.in},
Vivek Kumar Tiwari$^a$
\footnote{vivekkrt@gmail.com},
Anurag Tripathi$^b$
\footnote{anurag@hri.res.in}
}\\
\end{center}
\vspace{10pt}
\begin{flushleft}
{\it
a)~~ Department of Physics, University of Allahabad, Allahabad 211002, India. \\
b)~~Regional Centre for Accelerator-based Particle Physics,\\
~~~~Harish-Chandra Research Institute,
 Chhatnag Road, Jhunsi, Allahabad 211019, India.\\
}
\end{flushleft}
\vspace{10pt}
\centerline{\bf Abstract}
The first results on next-to-leading order QCD corrections to production of two $Z$ bosons  
in hadronic collisions in the extra dimension model of Randall and Sundrum are presented.
Various kinematical distributions are obtained to order $\alpha_s$ in QCD by 
taking into account all the parton level subprocesses.  
We estimate the impact of the QCD corrections on various observables and find that they
are significant.  We also show the reduction in factorization scale uncertainty when
${\cal O}(\alpha_s)$ effects are included.

\vskip12pt
\vfill
\clearpage

\setcounter{page}{1}
\pagestyle{plain}

\section{Introduction}
\newcommand{\vn}{\ensuremath{{\vec n}}}
The last missing piece of the standard model (SM), the Higgs boson,
remains elusive to this date, and it is hoped that the Large Hadron 
Collider ( LHC ) will shed light on the mechanism of spontaneous
symmetry breaking and discover the Higgs bosons. Even if it is discovered
there remain fundamental issues, such as the hierarchy problem and others,
which make us believe in the existence of some new physics beyond the 
standard model.
The LHC which will operate at an enormous centre of mass 
energy (${\sqrt S}=14~ \rm{TeV}$) promises
to be a discovery machine and it is hoped that some signals of 
new physics beyond the SM will be observed.
Extra dimension models 
\cite{Antoniadis:1998ig, ArkaniHamed:1998rs, 
Randall:1999ee, Randall:1999vf}
offer an attractive alternative to the 
supersymmetry to address the hierarchy problem.
In this paper we will consider
the extra dimension model of Randall and Sundrum (RS) 
\cite{Randall:1999ee, Randall:1999vf}.
There are many important discovery channels at the LHC
such as $\gamma \gamma, ZZ, W^+W^-$, jet production.  
In the SM the production of two $Z$ bosons is suppressed as it begins at the 
order $e^4$ in the the electromagnetic coupling and also because
of the large $ZZ$ production threshold.  
The two $Z$ bosons can couple to Kaluza Klein $(KK)$ gravitons, thus $ZZ$ pairs can be 
produced through virtual graviton exchange at the leading order. These
observations make $ZZ$ production one of the important discovery channels.

At hadron colliders Quantum Chromodynamics (QCD) plays an important role as
the {\it incoming} states in any scattering event are the partons, which
are described by parton distribution functions (${\rm pdf^s}$). The ${\rm pdf^s}$
depend on the factorization scale ($\mu_F$) which is, to a large extent, arbitrary.
This scale $\mu_F$ enters into any observable and makes it sensitive to the choice
of its value and any leading order computation suffers from this sensitivity. However,
as a computation beyond the leading order (LO) is carried out, the $\mu_F$ dependence
partially cancels yielding results less sensitive to the factorization scale. It 
also improves upon LO results in that it includes missing higher orders terms
of the perturbation series which can be large. 
It is, thus, the motivation of this 
paper to consider production of $Z$ boson pairs at the LHC 
at next-to-leading order (NLO) accuracy in the strong coupling constant in 
RS model.

Leading order studies for $ZZ$ production in the SM can be found
in \cite{Brown:1978mq}.  $Z$ pair with a large transverse momentum jet 
at LO was studied in \cite{Baur:1988cq}.
LO study for $ZZ$ production in the context of extra dimension model of RS 
was carried out in \cite{Park:2001vk} and coupling of radion
with gluon and top quark loop was cosidered in \cite{Das:2005na}.
Because of its importance $ZZ$ production has also been studied to NLO accuracy
in the SM \cite{Ohnemus:1990za ,Mele:1990bq, Jager:2006cp}.
These results were subsequently updated in \cite{Campbell:1999ah ,Dixon:1999di}.
NLO studies in  SM via gluon fusion were carried out in  
\cite{Glover:1988rg, Binoth:2008pr}.
These studies provide the precise estimate of higher order effects through $K$ factor 
as well as the sensitivity of the predictions to factorization scale.  Importantly,
the corrections turned out to be larger than the expectations based on
soft gluon effects justifying a full-fledged NLO computation taking into
all the processes. 
We presented NLO results for $ZZ$ production at the LHC in large extra dimension
model of Arkani-Hamed, Dimopoulos and Dvali
\cite{Antoniadis:1998ig, ArkaniHamed:1998rs} in \cite{Agarwal:2009xr} where it
was shown that the $K$ factors are large. 
Although NLO results are available in SM and ADD model they do not exist in literature in 
the context of RS model which is the material of the present paper.

The results which are presented in this paper are obtained using our 
NLO Monte Carlo code (which is implemented on FORTRAN 77) that can easily
accommodate any cuts on the final state particles and obtain various 
kinematical distributions. Our code is based on the method of two cutoff
phase space slicing (for a review of the method see \cite{Harris:2001sx}) 
to deal with soft and
collinear singularities in the real emission contributions.
We will use the matrix elements presented in 
\cite{Agarwal:2009xr} and refer the reader to this paper for further details.

In what follows we will first briefly describe the RS model 
and then present the numerical results and finally conclude.

\section{RS Model}
In the RS model
the single extra dimension $\phi$ is compactified on a ${\bf
S}^1/{\bf Z}^2$ orbifold with a radius $R_c$ which is
somewhat larger than the Planck length.
Two 3-branes, the Planck brane and the TeV brane, are located
at the orbifold fixed points $\phi=0,\ \pi$, with the
SM fields localized on the TeV brane.
The five-dimensional metric, which is {\it non-factorizable} or $warped$
is of the form
\begin{equation}
ds^2 = e^{-2{\cal K}R_c|\phi|}\eta_{\mu\nu}dx^{\mu}dx^{\nu}~+~R_c^2d\phi^2  
\label{eq1}
\end{equation}
where $0 \leq \phi \leq \pi$.
The the huge ratio
$\frac {M_{Pl}}{M_{EW}} \sim 10^{15}$ can be generated by the exponent
${\cal K}R_c$ which needs to be only of
${\cal O}(10)$ thereby providing a way of avoiding the hierarchy problem.
It was shown in \cite{Goldberger:1999uk, Csaki:1999mp}
that $R_c$ can be stabilized against quantum fluctuations
by introducing an extra scalar field in the bulk. 

The tower of massive Kaluza-Klein (KK) excitations of the graviton,
$h^{({n})}_{\mu\nu}$, interact with the SM particles by:
\begin{eqnarray}
{\cal L}_{int} & \sim & -{1\over {\overline{M_{Pl}}}}
T^{\mu\nu}(x) h^{(0)}_{\mu\nu}(x)
-{e^{\pi {\cal K} R_c} \over {\overline{M_{Pl}}}} \sum_{n=1}^{\infty}
T^{\mu\nu}(x) h^{(n)}_{\mu\nu}(x) \ .
\label{eq2}
\end{eqnarray}
$T^{\mu\nu}$ is the
symmetric energy-momentum tensor for the SM particles on the
3-brane, and ${\overline{M_{Pl}}}$ is the reduced Planck scale. 
The masses of the $h^{({n})}_{\mu\nu}$ are given by
\begin{eqnarray}
M_n & = & x_n {\cal K} ~e^{-\pi {\cal K} R_c} \ ,
\label{eq3}
\end{eqnarray}
where the $x_n$ are the zeros of the Bessel function $J_1(x)$.
The zero-mode couples weakly and decouples
but the couplings of the
massive RS gravitons are enhanced by the exponential $e^{\pi {\cal K}
R_c}$ leading to interactions of electroweak strength.
Consequently, except for the overall warp factor in the RS case,
the Feynman rules in the RS model are the same as those for the ADD case
\cite{Giudice:1998ck, Han:1998sg}

The basic parameters of the RS model are
\begin{eqnarray}
m_0 & = & {\cal K} e^{-\pi {\cal K} R_c} \ , \nonumber \\
c_0 & = & {\cal K}/{\overline{M_{Pl}}} \ ,
\label{eq4}
\end{eqnarray}
where $m_0$ is a scale of the dimension of mass and sets the scale for
the masses of the KK excitations, and $c_0$ is
an effective coupling. The interaction of massive KK gravitons with
the SM fields can be written as
\begin{equation}
{\cal L}_{int}  \sim
- {c_0 \over m_0} \sum_n^{\infty} T^{\mu\nu}(x)
 h^{(n)}_{\mu\nu}(x) \ .
\label{eq5}
\end{equation}
Since ${\cal K}$ is related to the curvature of the fifth dimension
we need to restrict it to small enough values to avoid effects
of strong curvature. On the other hand ${\cal K}$ should not
be too small compared to ${\overline{M_{Pl}}}$ because that would reintroduce
hierarchy. These considerations suggest $0.01 \le c_0 \le 0.1$.
For our analysis we choose to work with the RS parameters $c_0$ and
$M_1$ the first excited mode of the graviton rather then $m_0$.

Let us define ${\cal D}(Q^2)$ as the sum of KK graviton propagators 
\begin{eqnarray}
{\cal D}(Q^2) &=& \sum_{n=1}^\infty \frac{1}{Q^2 - M_n^2 + i M_n \Gamma_n}
\equiv {\lambda \over m_0^2} \ ,
\label{eq15}
\end{eqnarray}
where $M_n$ are the masses of the individual resonances ( see Eq.~\ref{eq3} ) and the $\Gamma_n$
are the corresponding widths.  The graviton widths are obtained by calculating
their decays into final states involving SM particles.  $\lambda$ is defined
as
\begin{eqnarray}
\lambda (x_s) & = & \sum_{n=1}^\infty
\frac{x_s^2 -x_n^2 -i \frac{\Gamma_n}{m_0} x_n}
     {x_s^2 -x_n^2 +  \frac{\Gamma_n}{m_0} x_n} \ ,
\label{eq16}
\end{eqnarray}
where $x_s=Q/m_0$.  We have to
sum over all the resonances to get the value of $\lambda(x_s)$. This is done
numerically and for a given value of $x_s$, we retain all resonances which
contribute with a significance greater than one per mil, and treat the
remaining KK modes as virtual particles (in which case the sum can be done
analytically).

As the gravitons couple to $Z$ bosons, $PP \rarrow ZZ$ can now also proceed through 
a process where  gravitons appear at the propagator level. These new channels 
make it possible to observe deviations from SM predictions if extra dimensions
exist. In the following we will consider spin-2 gravitons only at the propagator level
and investigate this process at NLO level.

\section{Numerical results}

In this section we present invariant mass ($Q$) and rapidity ($Y$) distribution of 
the $Z$ boson pairs. These kinematical variables are defined as
\be
Q^2 = (p_{Z_1} +p_{Z_2})^2, \quad \quad Y= \frac{1}{2} \ln \frac{P_1 \cdot q}{P_2 \cdot q},
\ee
where $P_1$ and $P_2$ are the momenta of colliding hadrons, and  $q=p_{Z_1} + p_{Z_2}$
denotes the sum of the $Z$-boson 4-momenta. In obtaining these distributions
all order $\alpha_s$ contributions have been taken into account. At leading order
in SM, the process proceeds through $q\qb$ initiated process. As the gravitons 
couple to the $Z$-bosons, $q\qb$ and $gg$ initiated processes with  virtual
gravitons also contribute at the same order in QCD. We have considered all the $q\qb, gg$ initiated 
one loop virtual and, $q\qb, qg, gg$ initiated real emission corrections to these processes, both 
in the SM and the gravity mediated processes,
and their interferences. 
At the virtual level we used method of Passarino and Veltman \cite{Passarino:1978jh}
to reduce tensor loop integrals to scalar integrals. In dealing with real emission
contributions we have used two cutoff phase space slicing method. Here, using two
small dimensionless slicing parameters $\delta_s$ and $\delta_c$, the singular (soft and collinear)
regions of phase space are separated from the finite hard noncollinear region. We will refer to the
sum of contributions to crosssection from virtual, soft and collinear regions as 2-$body$ contribution,
and from hard noncollinear region as 3-$body$ contribution. The soft singularities cancel between real
and virtual contributions and the collinear singularities were removed by mass factorization in 
${\overline{MS}}$ scheme, this gives the finite 2-$body$ contribution. Finally the kinematical distributions
were obtained by integrating the 2-$body$, 3-$body$ and leading order pieces over the phase space using
monte carlo methods. Individually 2-$body$ and 3-$body$ contributions depend on the slicing parameters 
$\delta_s$ and $\delta_c$ but the sum is invariant against variations of these parameters over a wide range.
For further analysis we will use $\delta_s=10^{-3}$ and $\delta_c=10^{-5}$. 
For further details please see \cite{Agarwal:2009xr}.

Below we present various distributions for the LHC with a center of mass energy of $14~TeV$ as a 
default choice. However we will also present some results for a center of mass energy of $10~TeV$ for the LHC.
For numerical evaluation,
the following SM parameters 
\cite{Amsler:2008zzb} are used
\be 
M_Z = 91.1876~ GeV,  \quad \sin^2 \theta_W = 0.231
\ee
where $\theta_W$ is the weak mixing angle.
For the electromagnetic coupling constant $\alpha$ we use $ \alpha^{-1} = 128.89$. 
CTEQ6 \cite{Pumplin:2002vw, Stump:2003yu} density sets are used for parton distribution 
functions. 2-loop running for the strong coupling constant has been used, and we have done 
calculation with 5 quark flavors and taken the masses of quarks to be zero.
The value of $\Lambda_{QCD}$ is
chosen as prescribed by the CTEQ6 density sets. At leading order  we
use CTEQ6L1 density set ( which uses the LO running $\alpha_s$ ) with the corresponding 
$\Lambda_{QCD}=165~MeV $. At NLO we use CTEQ6M density set ( which uses 2-loop running $\alpha_s$ )
with the $\Lambda_{QCD}=226~MeV $; this value of $\Lambda_{QCD}$ enters into the evaluation of the 
2-loop strong coupling.
The  default choice for the renormalization and factorization scale is the identification
to the invariant mass of the $Z$ boson pair ie., $\mu_F =\mu_R =Q$. Furthermore the
$Z$ bosons will be constrained to satisfy $|y_Z| < 2.5$, where $y_Z$ is 
the rapidity of a final state $Z$ boson .

\begin{figure}[ht]
\centerline{\epsfig{file=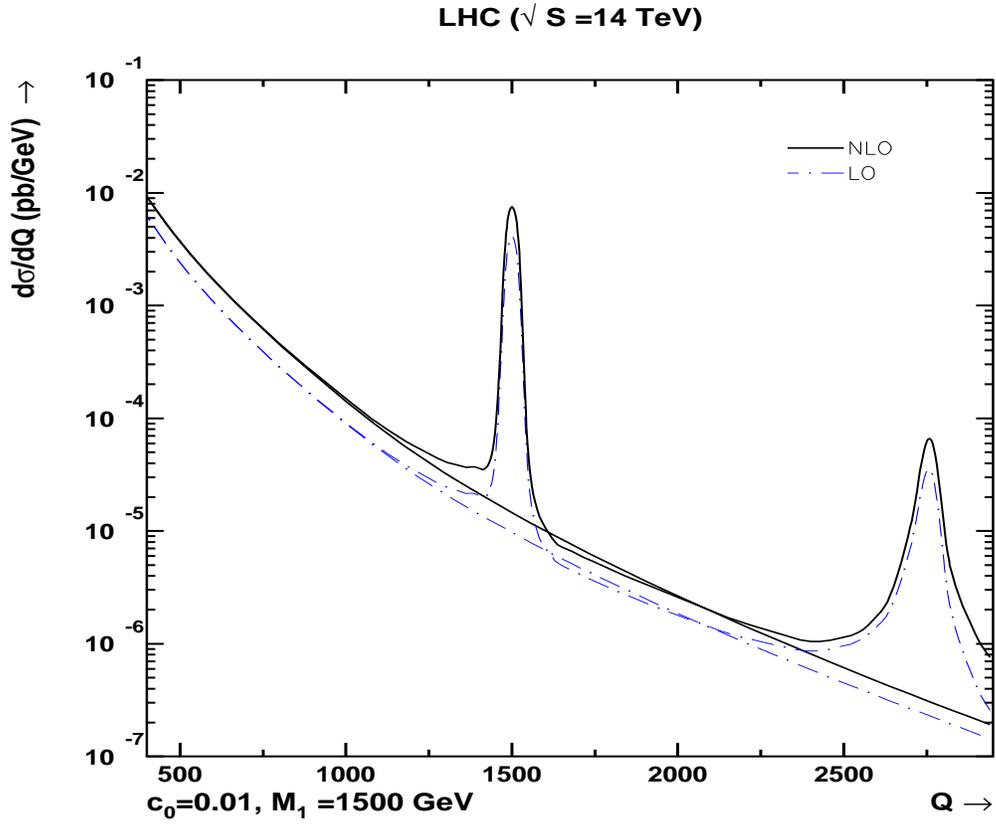,width=15cm,height=12cm,angle=0} }
\caption{Invariant mass distribution for SM and signal both at LO
and NLO. Dash-dot curves represent LO results and solid curves give
NLO results. We have chosen $M_1 =1500~GeV$ and the parameter $c_0 =0.01$.  }
\label{inv}
\end{figure}

\begin{figure}[ht]
\centerline{\epsfig{file=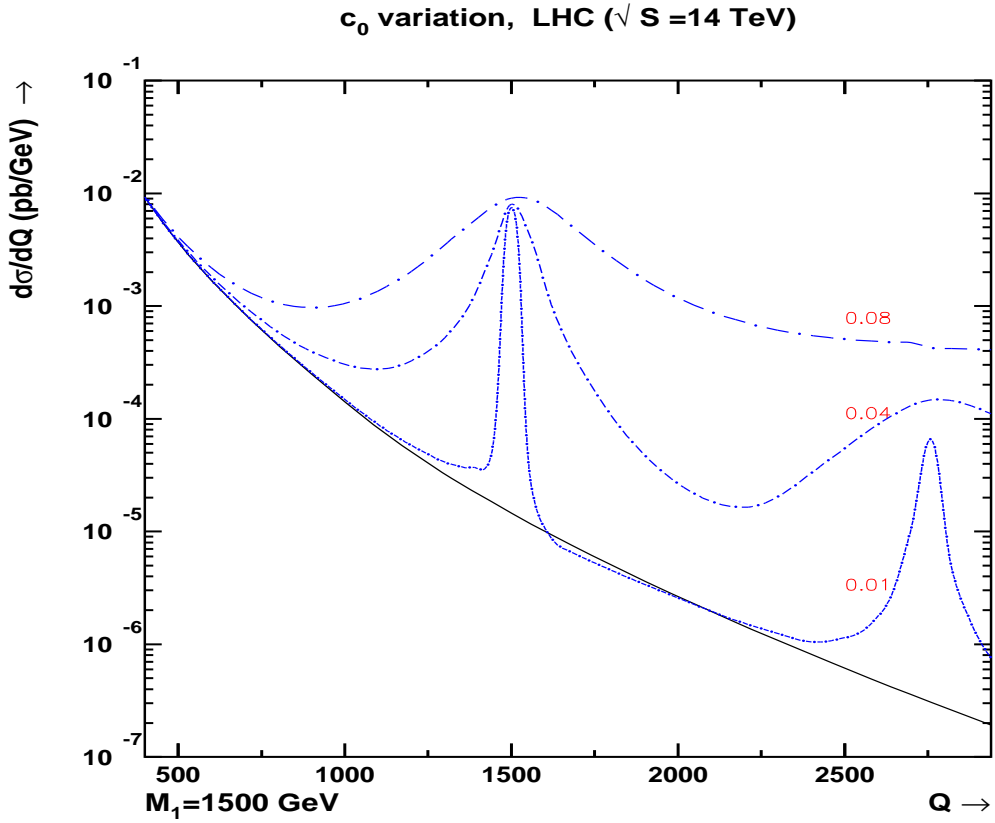,width=15cm,height=12cm,angle=0} }
\caption{
Effect of variation of $c_0$ on invariant mass distribution. 
All the curves correspond to NLO results with $M_1$ fixed
at $1500~GeV$. The solid curve corresponds to SM and the dash-dot curves
to the signal. The signal is plotted for $c_0 =0.01, 0.04, 0.08$ and 
the dash size increases with increasing $c_0$ }
\label{cvar}
\end{figure}

In Fig.~\ref{inv} we have plotted the invariant mass distribution both for the SM and the 
signal for LHC at $14 TeV$.
The two curves with peaks correspond to the signal and the remaining two curves give SM
predictions.  Here we have chosen $c_0 =0.01$ and $M_1 =1500 GeV$.
To highlight the importance of
QCD corrections we have also displayed the LO results of SM and the signal, and 
we observe that the $K$ factors (defined as $K=d\sigma^{NLO}/d\sigma^{LO}$) are large.
For the signal the $K$ factor is 1.82 at $Q=1500~GeV$. 
Also note that the gravitons appear as ever widening peaks, these resonance peaks are
clear signals of the RS model as opposed to the enhancement of the tail in ADD model.
Next we present in Fig.~\ref{cvar} the effects of varying the parameter $c_0$ on the
invariant mass distribution. All the curves shown correspond to NLO results, and
we have also plotted the SM background for comparison.

\begin{figure}[ht]
\centerline{\epsfig{file=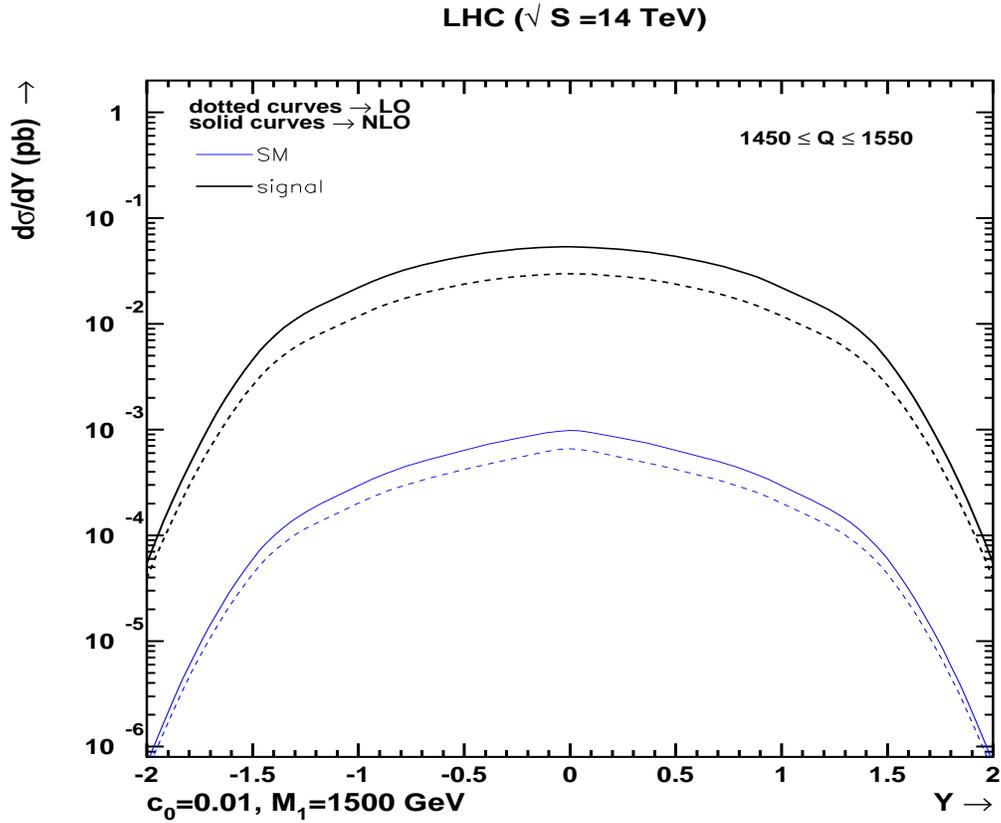,width=15cm,height=12cm,angle=0} }
\caption{Rapidity distribution for SM and signal both at LO
and NLO. Dash curves represent LO results and solid curves give
NLO results. We have chosen $M_1 =1500~GeV$ and the parameter $c_0 =0.01$.
To enhance the signal we have integrated over $Q$ in the range $1450 \leq Q \leq 1550$.  }
\label{y}
\end{figure}

\begin{figure}[ht]
\centerline{\epsfig{file=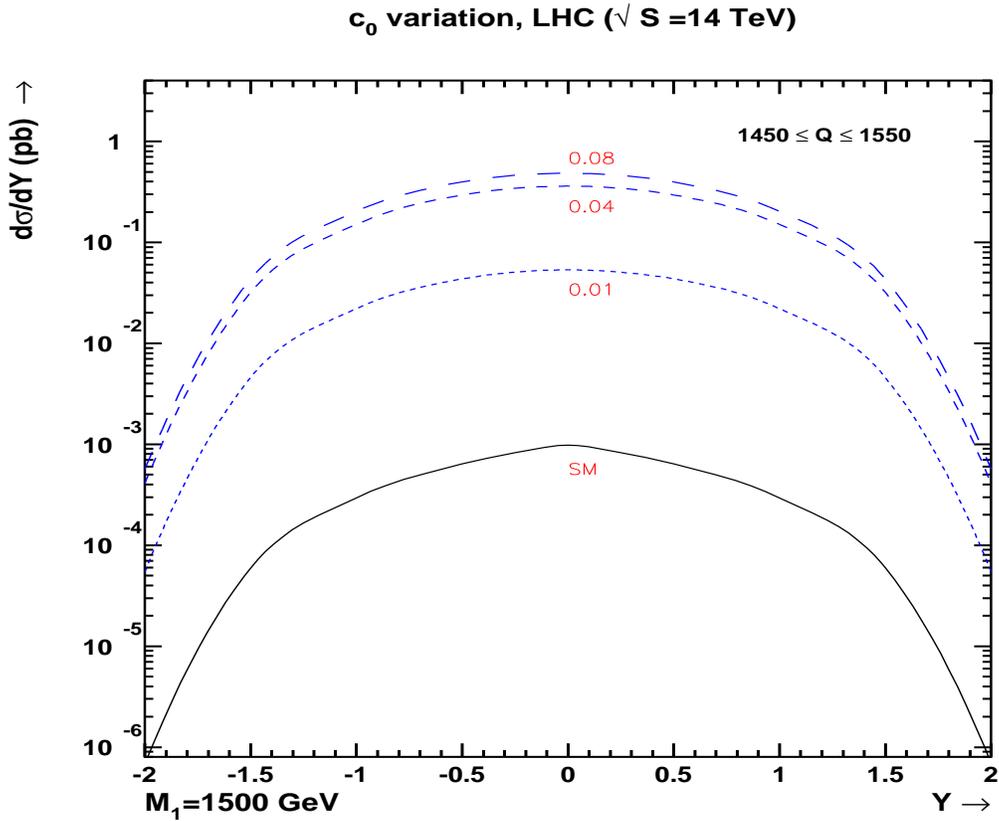,width=15cm,height=12cm,angle=0} }
\caption{Rapidity distribution for SM and signal at
NLO. Dash curves represent signal and solid curves gives SM contribution.
We have chosen $M_1 =1500~GeV$ and plotted the signal for $c_0 =0.01,~ 0.04,~ 0.08$.
To enhance the signal we have integrated over $Q$ in the range $1450 \leq Q \leq 1550$.  }
\label{yvar}
\end{figure}

In Fig.~\ref{y} we have plotted the rapidity distribution $d\sigma/dY $ at LO and 
NLO both for SM and the signal for $c_0=0.01$. 
We have plotted this distribution in the interval $-2.0 < Y < 2.0 $ and 
have carried out an integration over the invariant mass interval $1450 < Q < 1550$
to increase the signal over the SM background.
As expected the distribution is symmetric about $Y=0$.
In Fig.~\ref{yvar} we have shown the variation of rapidity distribution with
$c_0$.

\begin{figure}[ht]
\centerline{\epsfig{file=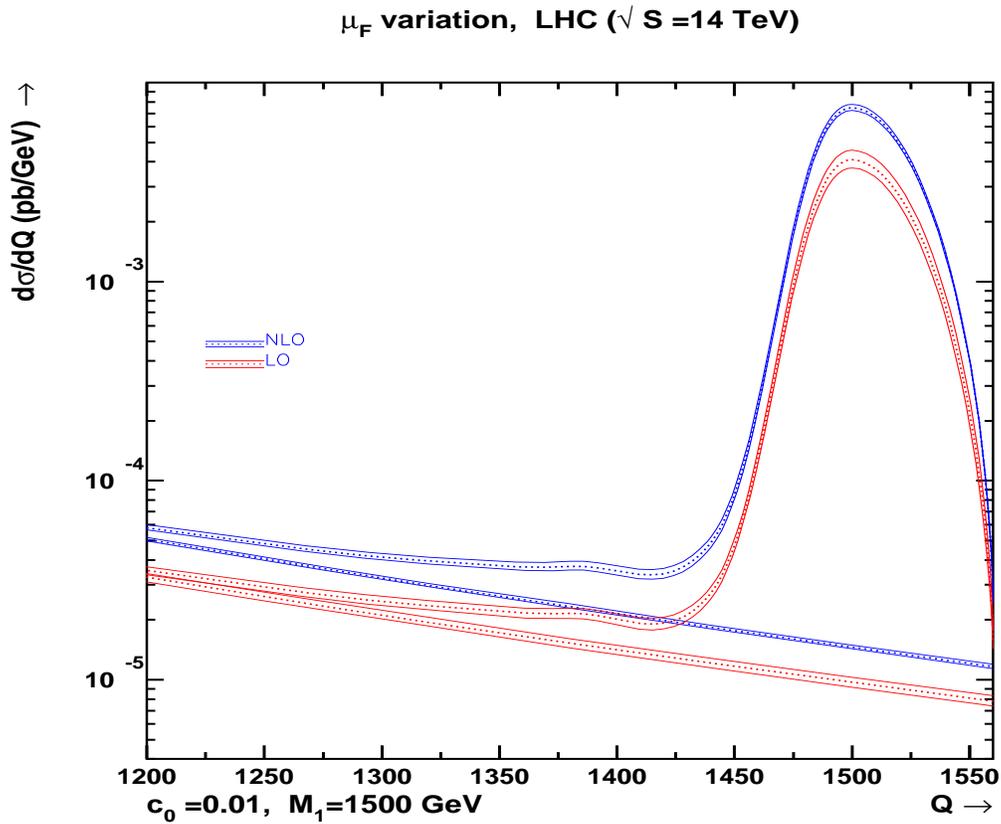 ,width=15cm,height=12cm,angle=0} }
\caption{Factorization scale variation in the invariant mass distribution.
The curves correspond to $c_0 =0.01$ and $M_1 =1500 GeV$ at the LHC at $\sqrt{S}=14~ TeV$.
The $\mu_F$ is varied between $Q/2$ and $2Q$. The dash curves correspond to $\mu_F=Q$  }
\label{mufvar}
\end{figure}

\begin{figure}[ht]
\centerline{\epsfig{file= 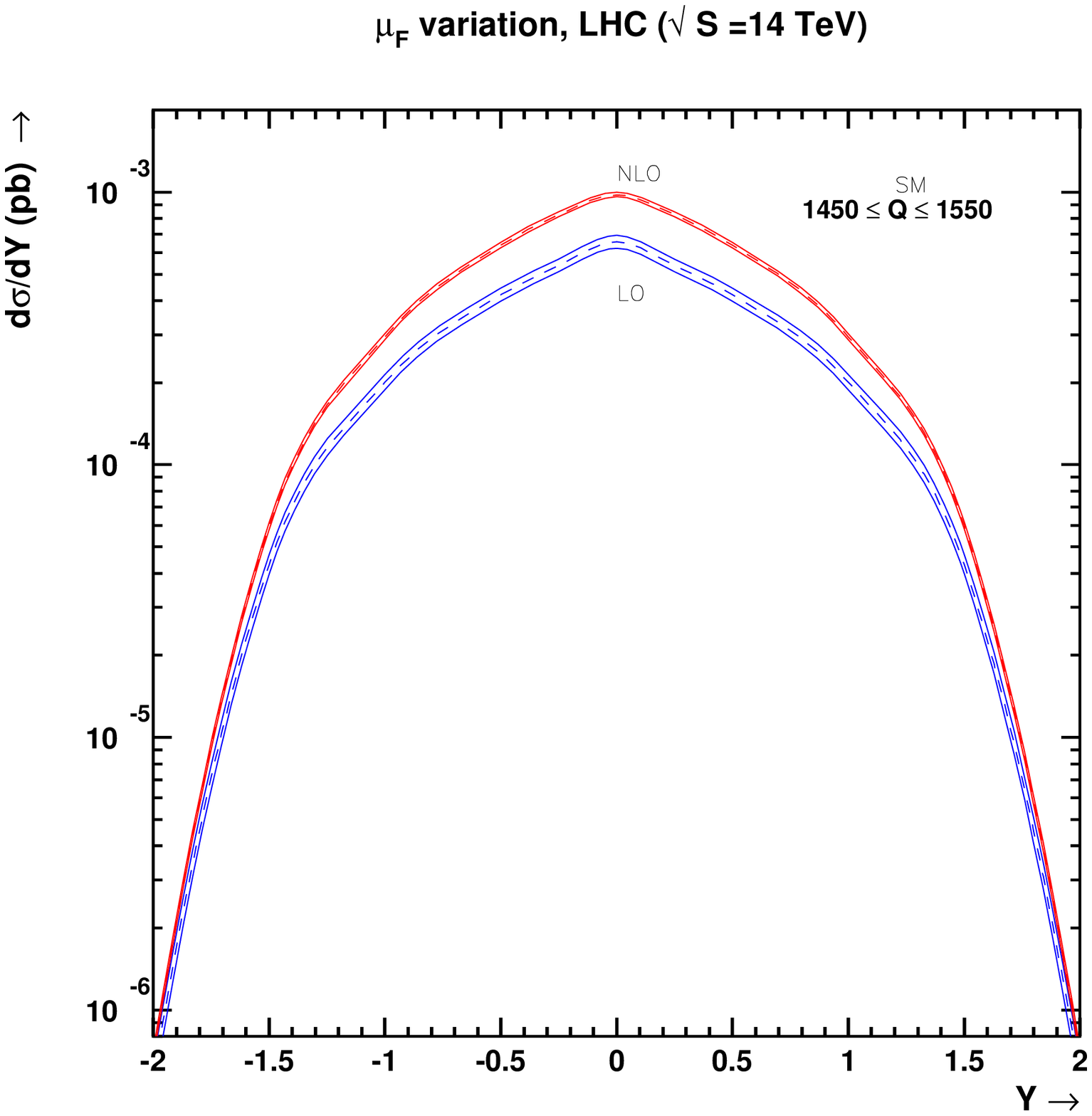 ,width=15cm,height=12cm,angle=0} }
\caption{Factorization scale variation in the rapidity distribution for the SM.
The curves correspond to $c_0 =0.01$ and $M_1 =1500 GeV$ at the LHC at $\sqrt{S}=14~ TeV$.
To enhance the signal we have integrated over $Q$ in the range $1450 \leq Q \leq 1550$.  
The $\mu_F$ is varied between $Q/2$ and $2Q$. The dash curves correspond to $\mu_F=Q$  }
\label{ysmmufvar}
\end{figure}

\begin{figure}[ht]
\centerline{\epsfig{file= 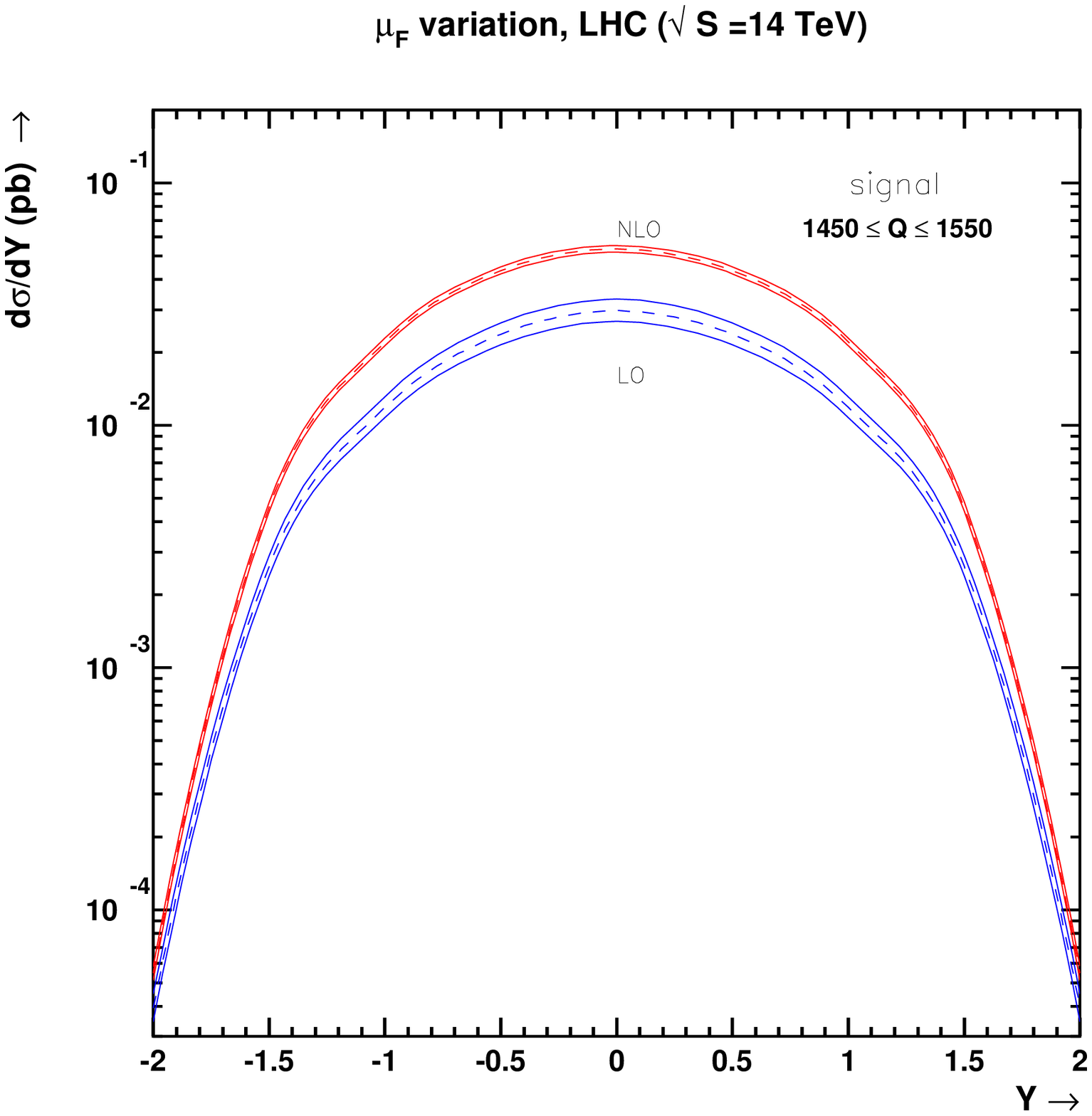 ,width=15cm,height=12cm,angle=0} }
\caption{Factorization scale variation in the rapidity distribution for signal.
The curves correspond to $c_0 =0.01$ and $M_1 =1500 GeV$ at the LHC at $\sqrt{S}=14~ TeV$.  
To enhance the signal we have integrated over $Q$ in the range $1450 \leq Q \leq 1550$.  
The $\mu_F$ is varied between $Q/2$ and $2Q$. The dash curves correspond to $\mu_F=Q$  }
\label{ysigmufvar}
\end{figure}

We have mentioned before that the NLO QCD corrections reduce the sensitivity 
of the cross sections to the factorization scale $\mu_F$; this we now show in 
the Fig.~\ref{mufvar}. We have plotted SM and the signal both at LO and NLO,
and have varied the factorization scale $\mu_F$ in the range $Q/2 < \mu_F < 2Q$.
The central curve in a given band (shown by the dotted curves) correspond to 
$\mu_F =Q$. In all these results the renormalization scale is fixed at $\mu_R =Q$.
We notice that the factorization scale uncertainty at NLO is much reduced compared
to the LO. 
For instance at $Q=1500~GeV$ varying $\mu_F$ between $Q/2$ to $2Q$
shows a variation of $20.6 \%$ at LO for the signal, however the NLO result at the same $Q$ value
shows a variation of $7.1 \%$. 
Similarly we show the dependence on factorization scale at LO and NLO in the 
rapidity distribution in Fig.~\ref{ysmmufvar} and Fig.~\ref{ysigmufvar} for
SM and signal respectively.

\begin{figure}[ht]
\centerline{\epsfig{file=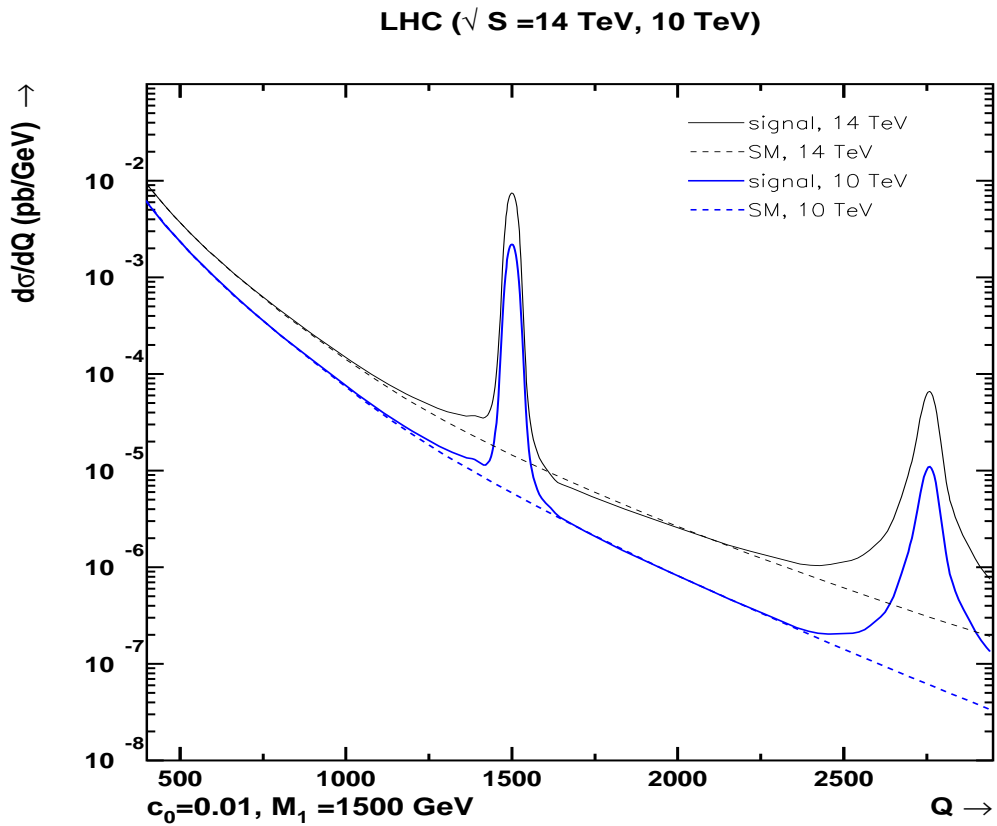,width=15cm,height=12cm,angle=0} }
\caption{Invariant mass distribution for SM and signal at $ \sqrt{S} =10 TeV$ and
$14 TeV$. All the curves correspond to NLO results. We have chosen $M_1 =1500~GeV$ and the parameter $c_0 =0.01$.  }
\label{ten}
\end{figure}

At the end we present in Fig.~\ref{ten}, $d\sigma/dQ$ for LHC with a centre of mass
energy of $10~TeV$ at NLO both for SM and signal. For comparison 
we have also plotted the $14~TeV$ results in the same figure.

\section{Conclusions}

In this paper we have carried out a full NLO QCD calculation for 
the production of two $Z$ bosons at the LHC at $14 TeV$ in the
extra dimension model of Randall and Sundrum. Here we take all order $\alpha_s$ contributions,
both in the SM and in the gravity mediated processes and their interferences,
into account. 
We have presented 
invariant mass and rapidity distributions both at LO and NLO. 
We use CTEQ 6L1 and CTEQ 6M parton density sets for LO and NLO observables, respectively.
Significant enhancements over the LO predictions are observed.
The $K$ factors are found to be large, for instance in invariant mass distribution 
the signal has a $K$ factor of 1.82 at $Q=1500~GeV$ (the position of first resonance). 
We have also presented the effects of variation of parameter $c_0$
both in $Q$ and $Y$ distributions. We have shown that a significant
reduction in LO theoretical uncertainty, arising from the factorization scale, is achieved 
by our NLO computation. 
For instance at $1500~GeV$ varying $\mu_F$ between $Q/2$ to $2Q$
shows a variation of $20.6 \%$ at LO for the signal, however the NLO result at the same $Q$ value
shows a variation of $7.1 \%$. 
Thus our NLO results are more precise than the LO results
and suitable for further studies for constraining the parameters of the RS model.
Invariant mass distribution is also presented for LHC at a center of mass energy of
$10 TeV$ at the NLO level.
\\

\noindent

\noindent
{\bf Acknowledgments:}
The work of NA is supported by CSIR Senior Research Fellowship, New Delhi.
NA, AT and VR would like to thank
the cluster computing facility at Harish-Chandra Research Institute. 
NA and VKT also acknowledge the computational support of the computing facility which has been developed by the Nuclear
Particle Physics Group of the Physics Department, Allahabad University under the Center of Advanced Study (CAS)
funding of U.G.C. India.
The authors
would like to thank Prakash Mathews and M.C. Kumar for useful discussions.

%

\end{document}